\documentclass{elsart}
\usepackage{graphics}
\usepackage{epsfig}
\usepackage{amssymb}
\begin{document}
\begin{frontmatter}
\title{Gamma-ray blazars: the view from AGILE}

\author{F.~D'Ammando\corauthref{cor}}$^{,~a}$
\corauth[cor]{Corresponding author}
%\thanks[footnote2]{Additional information regarding the corresponding author}
\ead{dammando@ifc.inaf.it}
\author{}
\author{A.~Bulgarelli$^{b}$, A.~W.~Chen$^{c}$, I.~Donnarumma$^{d}$, A.~Giuliani$^{c}$, F.~Longo$^{e}$,}
\author{L.~Pacciani$^{d}$, G.~Pucella$^{f}$, E.~Striani$^{d}$, M.~Tavani$^{d}$, S.~Vercellone$^{a}$,}
\author{V.~Vittorini$^{d}$ on behalf of the AGILE Team} 

\author{~~~~~~~~~~~~~~~~~~~~~~~~~~~~~~~~~~~~~~~and}
\author{S.~Covino$^{g}$, H.~A.~Krimm$^{h}$, C.~M.~Raiteri$^{i}$, P.~Romano$^{a}$, M.~Villata$^{i}$}

\address{$^{a}$INAF-IASF Palermo, Via Ugo La Malfa 153, I-90146 Palermo, Italy}
\address{$^{b}$INAF-IASF Bologna, Via Gobetti 101, I-40129 Bologna, Italy}
\address{$^{c}$INAF-IASF Milano, Via E. Bassini 15, I-20133 Milano, Italy}
\address{$^{d}$INAF-IASF Roma, Via Fosso del Cavaliere 100, I-00133 Roma, Italy}
\address{$^{e}$Dip. di Fisica and INFN, Via Valerio 2, I-34127 Trieste, Italy}
\address{$^{f}$ENEA Frascati, Vie E. Fermi 45, I-00044 Frascati (Roma), Italy}
\address{$^{g}$INAF, Oss. Astronomico di Brera, Via Bianchi 46, I-23807 Merate (LC),
  Italy}
\address{$^{h}$NASA/Goddard Space Flight Center, Greenbelt, MD 20771, USA}
\address{$^{i}$INAF, Oss. Astronomico di Torino, Via Osservatorio 20, I-10025 Pino
  Torinese (TO), Italy}

\begin{abstract}

During the first 3 years of operation the Gamma-Ray Imaging Detector onboard
the AGILE satellite detected several blazars in a high $\gamma$-ray
activity: 3C 279, 3C 454.3, PKS 1510$-$089, S5 0716$+$714, 3C 273, W
Comae, Mrk 421, PKS 0537$-$441 and 4C $+$21.35. Thanks to the rapid
dissemination of our alerts, we were able to obtain multiwavelength data from other
observatories such as \textit{Spitzer}, \textit{Swift}, RXTE, \textit{Suzaku},
INTEGRAL, MAGIC, VERITAS, and ARGO as well as radio-to-optical coverage by means of the
GASP Project of the WEBT and the REM Telescope.
This large multifrequency coverage gave us the opportunity to study
the variability correlations between the emission at different frequencies
and to obtain simultaneous spectral energy distributions of these sources
from radio to $\gamma$-ray energy bands, investigating the different
mechanisms responsible for their emission and uncovering in some cases
a more complex behaviour with respect to the standard models. We
present a review of the most interesting AGILE results on these
$\gamma$-ray blazars and their multifrequency data.
\end{abstract}

\begin{keyword}
Gamma-ray sources \sep quasars:active galactic nuclei \sep BL Lac objects \sep blazars
\PACS 98.70.Rz \sep 98.54.Cm 
\end{keyword}

\end{frontmatter}

\parindent=0.5 cm

\section{Introduction}

Gamma-ray astrophysics is an exciting field of astronomical sciences that
received a strong impulse in recent years. Detecting $\gamma$-ray emission in
the energy range from a few tens of MeV to a few tens of GeV is possible only
from space instrumentation, and in the past 20 years several space missions
had to face with the challenge of exploring the $\gamma$-ray domain. After the first
observations with the second NASA Small Astronomy Satellite (SAS-2; Fichtel et
al.~1975) and the European Cosmic ray Satellite (COS-B; Bennett 1990), the \textit{Compton
  Gamma-Ray Observatory (CGRO)} in the 1990s substantially increased our knowledge of
the $\gamma$-ray Universe and provided a wealth of data on a large variety of
sources as well as unsolved puzzles. In particular, the Energetic Gamma-Ray
Experiment Telescope (EGRET) onboard \textit{CGRO} operating in
the energy range 30 MeV--30 GeV carried out a complete sky survey
detecting hundreds of $\gamma$-ray sources (Fichtel et al.~1997; Hartman et
al.~1999). This scientific inheritance is the starting point for any high-energy astrophysics mission.

In particular, the detection of $\gamma$-ray loud Active Galactic Nuclei (AGNs) dates back to
the dawn of $\gamma$-ray astronomy, when COS-B detected
photons in the 50--500 MeV range from 3C 273 (Swanenburg et
al.~1978). However, 3C 273 remained the only AGN detected by COS-B. The
discovery of emission in the $\gamma$-ray domain from many AGNs by EGRET and by on-ground Cherenkov Telescopes was one of the most relevant breakthroughs of high energy
astrophysics in the last 20 years, leading to the identification of the first
class of $\gamma$-ray AGNs: the blazars (Punch et al.~1992; Hartman et al.~1999).

Blazars constitute the most extreme subclass of AGNs, characterized by the
emission of strong non-thermal radiation across the entire electromagnetic spectrum and in particular intense $\gamma$-ray emission above 100 MeV, dominating the extragalactic high-energy
sky. The typical observational properties of blazars include
irregular, rapid and often very large variability, apparent super-luminal
motion, flat radio spectrum, high and variable polarization at radio and
optical frequencies. These features are interpreted as resulting from the
emission of high-energy particles accelerated within a relativistic jet
closely aligned with the line of sight and launched in the vicinity of the
supermassive black hole (SMBH) harbored by the active galaxy (Blandford $\&$
Rees 1978; Urry and Padovani 1995). 

Blazars emit radiation across several decades of energy, from radio to TeV energy bands,
 and thus they are the perfect candidates for simultaneous observations at
 different wavelengths. EGRET observations together with ground-based
 Cherenkov Telescopes and coordinated multiwavelength observations provided
 the first evidence that the Spectral Energy Distributions (SEDs) of blazars
 are typically double humped (see e.g.~Mrk 421, Macomb et al.~1995), with the first peak occurring in the
IR/optical band in the so-called \textit{red blazars} (including Flat Spectrum Radio
Quasars, FSRQs, and Low-energy peaked BL Lacs, LBLs) and in UV/X-rays in the
so-called \textit{blue blazars} (including High-energy peaked BL Lacs, HBLs). The first peak is interpreted as synchrotron radiation from high-energy electrons in a
relativistic jet. The SED second component, peaking at MeV--GeV energies in
\textit{red blazars} and at TeV energies in \textit{blue blazars}, is commonly
interpreted as inverse Compton (IC) scattering of seed photons, internal or external to the jet, by highly relativistic
electrons (Ulrich et al.~1997), although other models involving hadronic
processes have been proposed (see e.g.~B\"ottcher 2007 for a recent review). 

Despite the large efforts devoted to the investigation of the radiation mechanisms
responsible for the high energy $\gamma$-ray emission in blazars the definitive
answer is still missing. The interest in blazars is now renewed thanks to the simultaneous presence of two $\gamma$-ray satellites, AGILE and
\textit{Fermi}, and the possibility to obtain $\gamma$-ray observations
over long timescales simultaneously with data collected from radio to TeV
energies, allowing us to reach a deeper insight on the jet structure and the
emission mechanisms at work in blazars. In this paper we present a review of
the AGILE results on the studies of $\gamma$-ray blazars.

\section{Blazars and AGILE}

The $\gamma$-ray observations of blazars are a key scientific project of
the \textit{Astrorivelatore Gamma ad Immagini LEggero} (AGILE) satellite
(Tavani et al.~2009). Thanks to the wide field of view of its
$\gamma$-ray imager ($\sim$ 2.5 sr), AGILE monitored tens of potentially
$\gamma$-ray emitting AGNs during each pointing. After November 2009
  AGILE, due to a reaction wheel failure, changed its scientific operation
  mode into a ``spinning mode'', with the instrument boresight axis sweeping
  the sky with the angular speed of 1\,$\deg$ s$^{-1}$. This allows the satellite to observe about 80$\%$ of the
  sky ever day, thus the number of $\gamma$-ray sources simultaneously monitored
by the satellite is still increased. In the first 3 years of operation, AGILE detected several blazars
during high $\gamma$-ray activity and extensive multiwavelength campaigns were
organized for many of them, providing the possibility to have long-term
observations of the brightest objects. The $\gamma$-ray activity timescales of
these blazars goes from a few days (e.g.~W Comae and 3C 273) to several weeks (e.g.~3C 454.3 and PKS 1510$-$089) and the flux
variability observed has been negligible (e.g.~3C 279 and Mrk 421), very rapid (e.g.~PKS
1510$-$089 and 3C 454.3), or extremely high (e.g.~3C 454.3 and 4C
$+$21.35). We note that at least one object for each blazar flavour (LBL, IBL,
HBL and FSRQ) was detected by AGILE, but only a few objects were detected more
than once in high activity and mainly already known strong $\gamma$-ray emitting sources. This evidence, together with the results
obtained by \textit{Fermi}-LAT during the first 11 months of operation (Abdo et
al.~2010b), suggests possible constraints on the properties of the most
intense $\gamma$-ray emitters. Recent studies in radio of a subsample of the
blazars detected by {\it Fermi}-LAT in the first three months showed that the
$\gamma$-ray emitters blazars have faster apparent jet speeds (Lister et al.~2009), wider apparent opening angles (Pushkarev et al.~2009), and
higher VLBI brightness temperatures (Kovalev et al.~2009) with respect to
the objects not detected in $\gamma$ rays. Future investigations
of a larger sample detected in $\gamma$ rays by \textit{Fermi} and AGILE could
give firm conclusion on the peculiar characteristics of the $\gamma$-ray blazars.

In the following we will present the most interesting results from the studies of the individual $\gamma$-ray blazars detected by AGILE.

\section{Individual Sources}

\subsection{3C 454.3}

Among the FSRQ class of blazars, 3C 454.3 is one of the brightest objects and also the source that exhibited
the most variable activity in the last years. In particular, during May 2005
the source was reported to undergo a very strong optical flare (Villata et
al.~2006). This exceptionally high state triggered observations by
high-energy satellites (RXTE: Remillard 2005; \textit{Chandra}: Villata et
al.~2006; INTEGRAL: Pian et al.~2006; \textit{Swift}: Giommi et al.~2006),
which confirmed an exceptionally high flux also in X-ray band. Unfortunately,
no $\gamma$-ray satellite was operative at that time. 

\begin{figure}[hhht!]
\begin{center}
\includegraphics[width=8.8cm]{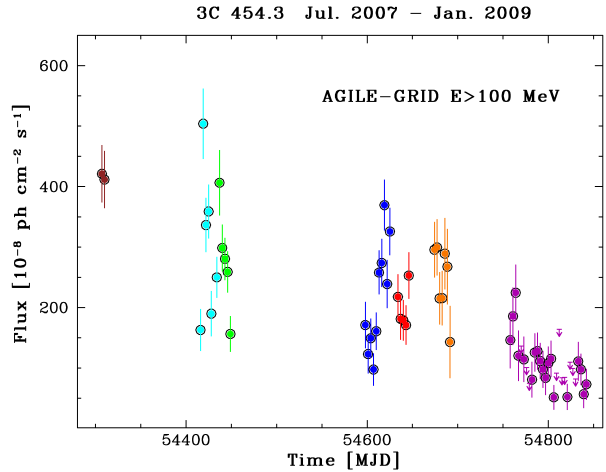}
\end{center}  
\caption{ AGILE GRID light curve of 3C 454.3 collected between July 2007 and
  January 2009 at $\sim$3 day resolution in
  units of 10$^{-8}$ ph cm$^{-2}$ s$^{-1}$ for E $>$ 100 MeV [Adapted from Vercellone et
  al.~2010].} 
\label{figure1}
 \end{figure}

In mid-July 2007, 3C~454.3 underwent a new optical brightening that triggered
observations at all frequencies, including a Target of Opportunity (ToO) by
the AGILE $\gamma$-ray satellite (Vercellone et al.~2008). That was the
beginning of an extraordinary long-term $\gamma$-ray activity of the source
until the huge $\gamma$-ray flares observed in early December 2009 (Striani et
al.~2010a) and April 2010 (Ackermann et al.~2010). In the period July 2007--January 2009 the
AGILE satellite monitored intensively 3C 454.3 together with the \textit{Spitzer},
GASP-WEBT, REM, MITSuME, \textit{Swift}, RXTE, \textit{Suzaku} and INTEGRAL
observatories, with two dedicated campaigns organized during November 2007 and
December 2007, as reported in Vercellone et al.~(2009) and
Donnarunna et al.~(2009b), respectively, and yielding the longest multiwavelength coverage of
this $\gamma$-ray quasar so far (Vercellone et al.~2010). 

\begin{figure}
\begin{center}
\includegraphics[width=9.1cm]{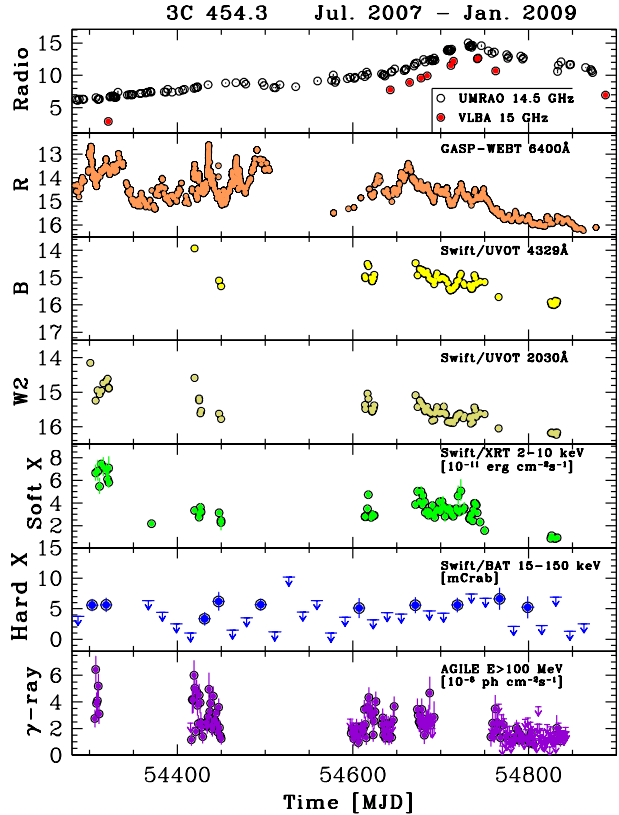}
\end{center}
  \caption{3C 454.3 light curves between July 2007 and January 2009 at
  increasing energies from top to bottom. Data were collected by AGILE,
  {\textit{Swift}} (BAT, XRT, and UVOT), GASP-WEBT, UMRAO, and VLBA [Adapted
  from Vercellone et al.~2010].} 
\label{figure2}
 \end{figure}

From the beginning of the AGILE operation the source underwent an
unprecedented long period of high $\gamma$-ray activity, playing the same role
for AGILE as 3C 279 did for EGRET (Hartman et al.~2001). The source showed flux levels variable on
short timescales of 24--48 hours, reaching a $\gamma$-ray flux higher than
500$\times$10$^{-8}$ ph cm$^{-2}$ s$^{-1}$ on daily timescales (see Fig.~\ref{figure1}, and Fig.~\ref{figure2} bottom panel). A diminishing trend of the $\gamma$-ray flux from July
2007 to January 2009 was observed with a hint of
a ``harder-when-brighter'' behaviour (see Vercellone et al.~2010, in particular
their Fig.~4),  previously observed in $\gamma$ rays only
for 3C 279 (Hartman et al.~2001) and marginally in PKS 0528$+$134 (Mukherjee
et al.~1999) in the EGRET \textit{era}. The optical
flux also appears extremely variable with a brightening of several tenths of
magnitude in a few hours (see Raiteri et al.~2008). Emission in the optical range
seems to be correlated with that at $\gamma$ rays, with a lag of the
$\gamma$-ray flux with respect to the optical one less than 1 day during
bright states. The correlation between the $\gamma$-ray flux and the optical
flux density during November--December 2007 was investigated (Vercellone et
al.~2010) by means of the discrete correlation function (DCF; Edelson $\&$ Krolik 1988). The
  corresponding DCF show a maximum DCF $\sim$ 0.38 for a null time lag. By calculating the centroid and estimating the uncertainty by means of the
  the ``flux randomization/random subset selection'' method (Peterson 2001)
  Vercellone et al.~found that the time lag between $\gamma$-ray and optical emission is -0.4$^{+0.6}_{-0.8}$
  days, i.e. about 10 hours. They obtained a similar
  time lag considering the period October 2008 -- January 2009, in agreement
  also with that obtained by Bonning et al.~(2009) by analyzing the public
  $\gamma$-ray data from {\it Fermi}-LAT and the optical SMARTS data. However,
superimposed to the overall trend some sub-structures on shorter timescales
with different variability could be present in the optical and $\gamma$-ray
bands (see e.g.~Tavecchio et al.~2010) .

From the comparison of the light curves from radio to $\gamma$ rays shown
in Fig.~\ref{figure2} it is noticeable that, while at almost all the
frequencies the flux shows a diminishing trend with time during the
period July 2007--January 2009, the 15 GHz radio core flux increases, although
no new jet component seems to be detected in the high resolution VLBA
images. The different behaviour observed in radio, optical and $\gamma$ rays
from the end of 2007 could be interpreted in the framework of a helical jet
model (see Villata et al.~2009) as the change in the jet geometry between 2007
and 2008. The change of orientation yields different alignment
configurations within the curved jet, therefore a different angle of view
with respect to the observer and consequently a different Doppler boosting of
the emission, as discussed in Vercellone et al.~(2010).   

\begin{figure}[hhht!]
\begin{center}
\includegraphics[width=8.8cm, angle=0]{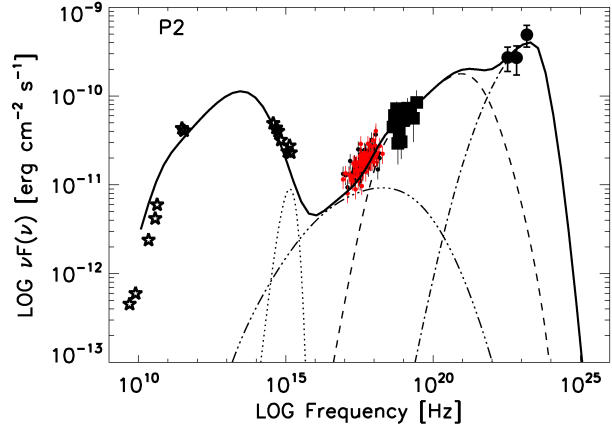}
\end{center}
  \caption{Spectral Energy Distribution of 3C 454.3 including AGILE/GRID,
   INTEGRAL/IBIS, {\it Swift}/XRT, {\it Swift}/UVOT, and GASP-WEBT data
   collected in the period 19--22 November 2007. The dotted, dashed, dot-dashed, and triple-dot-dashed lines represent
   the accretion disk black body, the external Compton on the disk, the external Compton
   on the BLR and the SSC radiation, respectively [Adapted from Vercellone et
  al.~2009].} 
\label{figure3}
 \end{figure}

Considering the wide coverage obtained over the entire electromagnetic
spectrum, we had the opportunity to build
time-resolved SED of 3C 454.3 at different epochs and study in detail the emission
mechanisms at work in this source.
As shown in Fig.~\ref{figure3}, the dominant emission mechanism above 100 MeV in 3C 454.3 seems to be the IC scattering of relativistic electrons in the jet on the
external photons from the Broad Line Region (BLR; see also Vercellone et
al.~2009,~2010), even if in some cases the contribution of external Compton (EC) of seed photons from a hot corona
(Donnarumma et al.~2009b) or an infrared dusty torus (Sikora et al.~2008) could
also be important. Moreover, the long-term monitoring confirmed the presence of an
important contribution of the accretion disk emission during the low activity
states of 3C 454.3 (Vercellone et al.~2010), as already detected by Raiteri et al.~(2007) with
simultaneous GASP and XMM-{\it Newton} observations.

\begin{figure}[hhht!]
\begin{center}
\includegraphics[width=8.8cm, angle=-90]{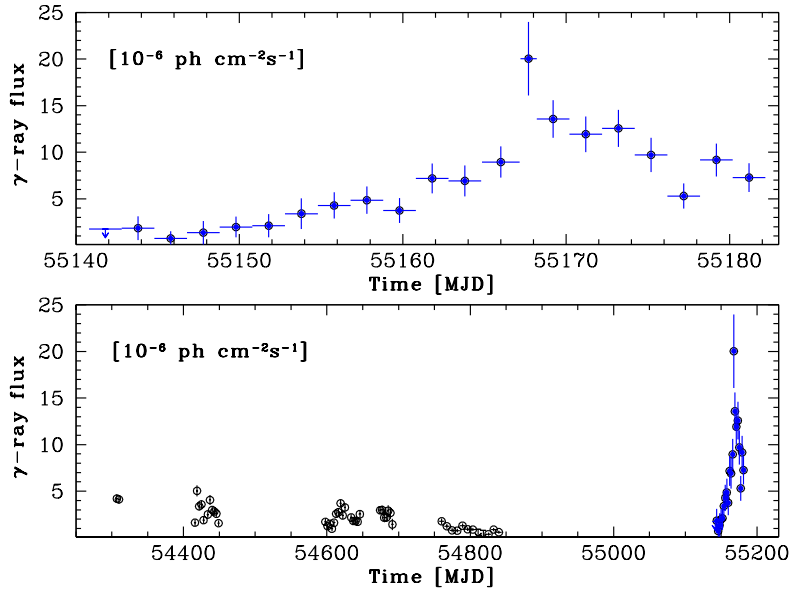}
\end{center}
  \caption{Gamma-ray light curves of 3C 454.3 as monitored by AGILE. The
  bottom panel shows the overall period between July 2007 and January 2010. The top panel is focused only on the
  period between 7 November 2009 and 9 January 2010 [Adapted from Striani et
  al.~2010a].} 
\label{figure4}
 \end{figure}

On 2--3 December 2009, 3C 454.3 became the brightest $\gamma$-ray
source in the sky, reaching a peak flux of about 2000 $\times$ 10$^{-8}$ ph
cm$^{-2}$ s$^{-1}$ (Fig.~\ref{figure4}). Intensive multifrequency
observations showed an overall correlation at all wavelengths for both long
and short timescales. However, the unusual $\gamma$-ray super-flaring activity
was not accompanied by strong emission of similar intensity in the optical or
even in the soft X-ray bands (see Pacciani et al.~2010). The pre- and post-flare broad band spectrum can
be adequately represented by a simple one-zone synchrotron self Compton (SSC)
model plus EC emission in which the accretion disk and the BLR provide the necessary
soft radiation field for the IC components. Instead, the spectrum of the 2--3 December 2009 super-flare would require with respect to the pre-flare an increase of the electron energy
and density, and a slight reduction of the comoving magnetic field for the whole
electron population of the blob (see Bonnoli et al.~2011). Pacciani et
al.~(2010) use a different approach, assuming a long-term rise and fall of the accretion rate onto the
central black hole that causes an overall increase of the synchrotron emission
and of the seed photons scattered by the primary component of accelerated
electrons. An additional population of electrons, due to a further particle acceleration and/or plasmoid ejection
near the jet basis, could be present during the super-flare. It is worth mentioning that, although rare, flaring episodes with
similar extreme energetics are not unique, neither in this object (see
e.g. Striani et al.~2010c, Sanchez and Escande 2010) nor in other FSRQ (see the $\gamma$-ray flare of 4C $+$21.35 in June 2010;
Striani et al.~2010b, Iafrate et al.~2010). 

\subsection{PKS 1510--089}

PKS 1510$-$089 is another blazar that in the last three years showed high
variability over the whole electromagnetic spectrum. In particular, high
$\gamma$-ray activity was observed by AGILE and \textit{Fermi}-LAT. AGILE
detected intense flaring episodes in August 2007 (Pucella et al.~2008) and
March 2008 (D'Ammando et al.~2009) and an extraordinary activity during
March 2009 (D'Ammando et al.~2010). 

During the period 1--16 March 2008, AGILE detected
an average $\gamma$-ray flux from PKS 1510$-$089 of (84 $\pm$ 17)$\times$10$^{-8}$ ph cm$^{-2}$ s$^{-1}$ for E $>$ 100 MeV. The flux measured between 17 and 21 March
was a factor of 2 higher, with a peak value of (281 $\pm$ 68)$\times$10$^{-8}$
ph cm$^{-2}$ s$^{-1}$ on 19 March 2008. Moreover, between
January and April 2008 the source showed an intense optical activity, with
several flaring episodes of fast variability. A significant increase of the
flux was observed also at submillimetric frequencies in mid-April, suggesting
that the mechanisms producing the flaring events in the optical and
$\gamma$-ray bands could also be responsible for the high activity
  observed in the sub-mm band, with a delay likely due to an opacity effect. 

The $\gamma$-ray flare triggered 3 \textit{Swift} ToO observations in
three consecutive days between 20 and 22 March 2008. The first XRT observation
showed a very hard X-ray photon index ($\Gamma$ = 1.16 $\pm$ 0.16) with a flux
in the 0.3--10 keV band of (1.22 $\pm$ 0.17) $\times$10$^{-11}$ erg cm$^{-2}$ s$^{-1}$ and a decrease of the flux of about 30$\%$ between 20 and
21 March. The \textit{Swift}/XRT observations showed a harder-when-brighter
behaviour of the spectrum in the X-ray band, confirming a behaviour already
observed in this source by Kataoka et al.~(2008), a trend usually observed in HBLs but quite rare in FSRQs such as PKS 1510$-$089. 

\begin{figure}[!hhhb]
\begin{center}
\includegraphics[width=8.8cm]{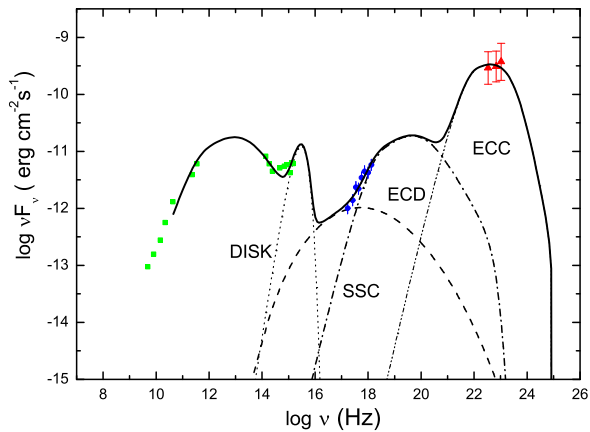}
\end{center}
  \caption{SED of PKS 1510$-$089 in mid-March 2008 with AGILE, {\textit{Swift}} and GASP-WEBT data. The dotted, dashed,
  dot-dashed, and double dot-dashed lines represent the accretion disk
  emission, the SSC, the external Compton on the disk radiation (ECD) and on
  the BLR radiation (ECC), respectively [Adapted from D'Ammando et al.~2009].} 
\label{figure5}
\end{figure}

\noindent This harder-when-brighter behaviour could be due to the
contamination at low energies of a second component with respect to the EC
emission that usually dominates the X-ray spectrum of the FSRQs. In PKS
1510$-$089 the X-ray spectral shape of the EC component should remain almost
constant. On the contrary, the contribution of the second component
should vary according to the change of the source activity level, being more
important when the source is fainter and almost hidden by the EC emission when
the source is brighter. Different origins were proposed for this second component: the soft
X-ray excess (Arnaud et al.~1985, Gierlinski $\&$ Done 2004), an important
contribution of the SSC component in any activity states or a feature of the bulk Comptonization (Celotti et al.~2007). No conclusive
evidence about its nature has been obtained so far. The SED of the AGILE
observation of 17--21 March 2008 and the simultaneous data collected from radio-to-X-rays by GASP-WEBT and \textit{Swift} is modeled with thermal emission
of the disk, SSC model plus the contribution by EC scattering of direct disk
radiation and of photons reprocessed by the BLR (see Fig.~\ref{figure5}). Some
features in the optical/UV spectrum clearly indicate the presence of Seyfert-like components, such as the little and big blue bumps.

\begin{figure}[!ht]
\begin{center}
\includegraphics[width=8.8cm]{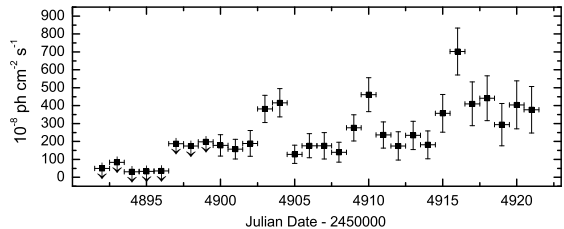}
\end{center}
  \caption{AGILE $\gamma$-ray light curve of PKS 1510$-$089 between 1 and 30
  March 2009 for E $>$ 100 MeV. The downward arrows
  represent 2-$\sigma$ upper limits [Adapted from D'Ammando et al.~2010].} 
\label{figure6}
 \end{figure}

Moreover, PKS 1510$-$089 showed an extraordinary $\gamma$-ray activity during March
2009, with several flaring episodes and a flux that reached
700$\times$10$^{-8}$ ph cm$^{-2}$ s$^{-1}$ (Fig.~\ref{figure6}). During
February--March 2009 the source also showed an increasing activity from
near-IR to UV, as monitored by GASP-WEBT, REM and \textit{Swift}/UVOT, with a
flaring episode on 26--27 March 2009, suggesting that a single mechanism is
responsible for the flux enhancement observed in the low-energy part of the
spectrum at the end of March. On the contrary, the
\textit{Swift}/XRT observations show no clear correlation of the X-ray
emission with the optical and $\gamma$-ray ones. These X-ray
observations, as well as the March 2008 observations, show a hard photon
index ($\Gamma_{\rm x}$ = 1.3--1.6) with respect to most FSRQs and a hint of harder-when-brighter behaviour due to the possible presence of a second
emission component in the soft X-ray part of the spectrum. This second component could be associated to the soft X-ray excess rather than with a SSC
contribution. During March 2009 two short flaring episodes were detected in the
15--50 keV energy band by \textit{Swift}/BAT: the first covered approximately 2 days,
beginning on 8 March with a flux of 28 mCrab and peaking on 9 March at 40
mCrab; the second on 29 March with a flux of 15 mCrab. It is noteworthy that the first hard X-ray outburst detected by BAT occurred just at the
beginning of the $\gamma$-ray activity observed by AGILE.

\begin{figure}[!hhhb]
\begin{center}
\includegraphics[width=8.8cm]{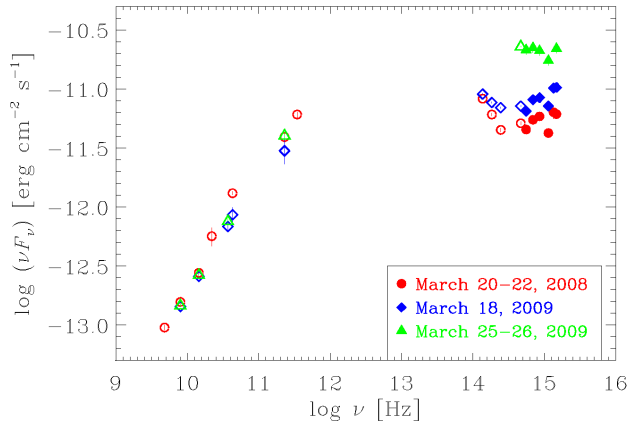}
\end{center}
  \caption{SED of the low-energy part of the spectrum of PKS 1510$-$089
  constructed with data collected by GASP-WEBT, REM and {\textit{Swift}}/UVOT
  during March 2008 and March 2009 [Adapted from D'Ammando et al.~2010].} 
\label{figure7} 
\end{figure}

In Fig.~\ref{figure7} we compare the SED
from radio-to-UV for 25--26 March 2009 with those collected on 20--22 March
2008 and 18 March 2009. The SED collected on 18
March 2009 confirmed the evidence of thermal signatures in the optical/UV
spectrum of PKS 1510--089 also during high $\gamma$-ray states. On the other
hand, the broad band spectrum from
radio-to-UV during 25--26 March 2009 show a flat spectrum in the optical/UV
energy band, suggesting an important contribution of the synchrotron emission
in this part of the spectrum during the brighter $\gamma$-ray flare and
therefore a significant shift of the synchrotron peak, usually observed in
this source in the infrared band. This is in agreement with the drastic change
of polarization angle observed in VLBA data simultaneously with the optical
flare (see Marscher et al.~2010). The increase of the synchrotron emission leads to
the decrease of the evidence of the thermal features observed in the other
SEDs. This indicates that the main contributor to the optical/UV emission in
this object could be alternatively a thermal or non-thermal mechanism during
the different activity periods (D'Ammando et al.~2010).

\subsection{S5 0716+714}

The intermediate BL Lac (IBL) object S5 0716+714 was observed by AGILE
during two different periods: 4--23 September and 22 October--1 November 2007, as
discussed in detail in Chen et al.~(2008). In particular, between 7 and 12 September 2007 the source showed a high $\gamma$-ray activity with an average
flux of F$_{\rm E > 100\, MeV}$ = (97 $\pm$ 15)$\times$10$^{-8}$ ph cm$^{-2}$ s$^{-1}$
and a daily peak of F$_{\rm E>100\, MeV}$ =
(193 $\pm$ 42)$\times$10$^{-8}$ ph cm$^{-2}$ s$^{-1}$, with an increase of flux by
a factor of four in three days. 
The $\gamma$-ray flux detected by AGILE
is the highest ever detected from this object and one of the highest fluxes observed from
a BL Lac object. The intense $\gamma$-ray flare detected by AGILE in mid-September 2007 triggered
optical observations by GASP-WEBT. S5 0716$+$714 brightened from $R$ = (12.92 $\pm$ 0.01) mag on 8
September to $R$ = (12.58 $\pm$ 0.04) mag on 12 September, and faded to $R$ $\sim$
13.01--13.03 mag on 15 September (Carosati et al.~2007). 

About one month later, the
GASP-WEBT observed a new bright phase of the source, triggering AGILE and {\it Swift} ToO observations. In particular, after a rather variable phase, the optical flux in mid-October
started to rise, reaching a peak
of $R$ = (12.15 $\pm$ 0.01) mag on 22 October. This is the highest optical brightness level ever observed from
this source. Moreover, a rare roughly contemporaneous radio-optical outburst seems to be detected by GASP-WEBT. However, when seen in detail, the event in the two
bands showed different behaviours: the optical flux presents stronger and faster
variations, whereas the radio flux rises and falls in a much smoother way (Villata et al.~2008).
Another very intense $\gamma$-ray flare, at a flux level of the order of 200$\times$10$^{-8}$ ph cm$^{-2}$ s$^{-1}$, was
detected by AGILE on 22--23 October 2007, simultaneously to the optical flare.
 
\begin{figure}[!hhht]
\begin{center}
\includegraphics[width=8.8cm]{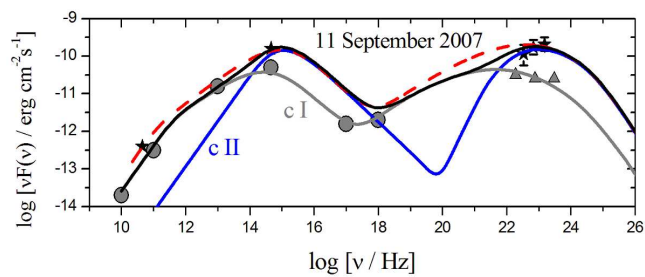}
\end{center}
  \caption{SED of S5 0716$+$714 on 11 September 2007 including optical
  GASP-WEBT and $\gamma$-ray AGILE data (black stars). Historical data
  relative to a ground state and EGRET data are represented with grey
  dots and triangles, respectively. Curves labelled as ``c I'' and ``c II'' represent the two separate
  components. The solid black line and the dashed red lines represent the
  two-component and one-component model, respectively [Adapted from
  Vittorini et al.~2009].}
\label{figure8} 
 \end{figure}

The SED built with the simultaneous data collected by AGILE and GASP-WEBT in
mid-September 2007 is consistent with a SSC model, but only when including
two SSC components of different variability. Together with the first SSC
component, that is slowly variable and reproduces the ground state, Chen et
al.~(2008) and Vittorini et al.~(2009) add a faster second SSC component dominating in the optical and $\gamma$-ray
bands (see Fig.~\ref{figure8}). 

Nilsson et al.~(2008) estimated the redshift of the source (z = 0.31
$\pm$ 0.08) from optical imaging of the underlying galaxy. This allowed us
to calculate the total power transported in the jet, which is extremely high for the two flaring episodes, approaching or slightly
exceeding the maximum power generated by a spinning BH of 10$^{9}$ M$_\odot$
through the pure Blandford-Znajek (BZ) mechanism with conservative values of the
 magnetic field B (Vittorini et al.~2009). S5 0716$+$714 is the first BL Lac object that approached the limit
of the BZ mechanism and eventually exceeded it. If confirmed, this
violation could be explained in terms of the alternative Blandford-Payne
mechanism (Blandford $\&$ Payne 1982) that, however, requires an ongoing accretion not
supported by the observations of S5 0716$+$714. Alternatively, such a high
power could be due to a less conservative value of the magnetic field, up
to $B^{2}$/4$\pi$ $\leq$ $\rho$c$^2$, related to
particle orbits plunging from the disk toward the BH horizon into a region
influenced by strong gravity effects (Meier 2002). 

After the $\gamma$-ray flaring episode of 22--23 October, AGILE observed the
source with a dedicated re-pointing between October 24 and November 1. The
source was detected at a $\gamma$-ray flux about a factor of 2
lower than the
September one with no significant variability. Simultaneously, \textit{Swift}
observed strong variability (up to a factor of $\sim$ 4) in soft X-rays, moderate
variability at optical/UV (less than a factor of 2), and approximately constant
hard X-ray flux. The different variability observed in optical/UV, soft
and hard X-rays is indicative of the injection of a second component (Giommi
et al.~2008), in agreement with the two SSC components used for modelling the SED
of the two $\gamma$-ray flares in September and October 2007.

\subsection{The Virgo Region: 3C 279 and 3C 273}

The Virgo region is one of the best studied regions of the sky by the \textit{CGRO}, especially with EGRET, but also with
OSSE and Comptel. During the \textit{CGRO} observations the presence of two bright and variable
$\gamma$-ray blazars was revealed: 3C 279 and 3C 273. Thus, the AGILE satellite performed
dedicated pointings of this region for investigating in detail the properties of these two blazars. 

3C 279 is the first extragalactic source detected by AGILE in mid-July 2007,
at the beginning of the Science Performance Verification Phase of the satellite,
as reported in Giuliani et al.~(2009). The average
$\gamma$-ray flux between 9 and 13 July 2007 is F$_{\rm E>100\, MeV}$ = (210 $\pm$
38)$\times$10$^{-8}$ ph cm$^{-2}$ s$^{-1}$, a flux level similar to the highest
one observed by EGRET and \textit{Fermi}-LAT. The simultaneous {\it
  Swift}/XRT observations performed between 10 and 13 July 2007 detected the
source with an unabsorbed 2--10 keV flux nearly constant at about 10$^{-11}$
erg cm$^{-2}$ s$^{-1}$ and a photon index of 1.4--1.5. No significant daily
variability was detected in the $\gamma$-ray flux of 3C 279 during the short
AGILE observation. Instead, the spectrum observed during the flaring episode
by AGILE is softer with respect to the previous EGRET observations. 

This could be an indication of a low accretion state of the
disk occurred some months before the $\gamma$-ray observations, suggesting a
dominant contribution of the external Compton scattering of direct disk 
radiation (ECD) compared to the external Compton scattering of the BLR clouds (ECC). As a matter of fact, a strong minimum in the optical band was
detected by the REM telescope two months before the AGILE observations. The reduction
of the activity of the disk should cause the decrease of the photon seed
population produced by the disk and subsequently a deficit of the ECC component with
respect to the ECD, an effect delayed by the light travel time required to the
photons to go from the inner disk to the BLR. EGRET observations of 3C 279
hinted at a gradual hardening of the spectrum during the flaring states that
can be interpreted as the ECC component dominating during the flaring
states (Hartman et al.~2001). Only one flare, during EGRET observation P9,
showed a soft ECD dominated spectrum. The AGILE observation seems to be similar to the P9 flare,
supporting the idea that a soft spectrum during flaring episodes is not a
extremely rare event. Only the long-term monitoring of this source in
$\gamma$ rays with the AGILE and {\it Fermi} satellites, together with the radio to
optical data, would clarify the exact nature of the seed photons for the IC
scattering and more generally provide us a new level of insight on the
jet structure and the emission mechanisms in this blazar. Recently, the
coincidence of a $\gamma$-ray flare of 3C 279 observed by \textit{Fermi}-LAT with the
dramatic change of optical polarization angle measured by the KANATA telescope
suggested
co-spatiality of the optical and $\gamma$-ray emission and provided evidence
for the presence of highly ordered magnetic fields and a non-axisymmetric
structure of the emission zone. This could imply a curved structure of the jet
and a dissipation region far from the central BH (Abdo et al.~2010c).

On the other hand, 3C 273 is a peculiar AGN that shows both properties
characteristic of a blazar, such as strong radio emission, apparent superluminal
jet motion, large flux variations and a two-humped SED (see Courvoisier et al.~1998 for a review), and other features typical of Seyfert
galaxies, such as a broad Fe emission line, the soft X-ray excess and the big blue
bump. Surprisingly, 3C 273 was discovered to emit in $\gamma$ rays by COS-B in
1976 (Swanenburg et al.~1978). EGRET pointed this FSRQ several times, not always
detecting it, with an average flux over all the EGRET observations of (15.4
$\pm$ 1.8)$\times$10$^{-8}$ ph cm$^{-2}$ s$^{-1}$ (E $>$ 100 MeV). Recently \textit{Fermi}-LAT detected two exceptional $\gamma$-ray outbursts by
3C 273, with a peak flux of $\sim$ 1000$\times$10$^{-8}$ ph cm$^{-2}$ s$^{-1}$
(Abdo et al. ~2010a).

\begin{figure}[hhht!]
\begin{center}
\includegraphics[width=8.8cm]{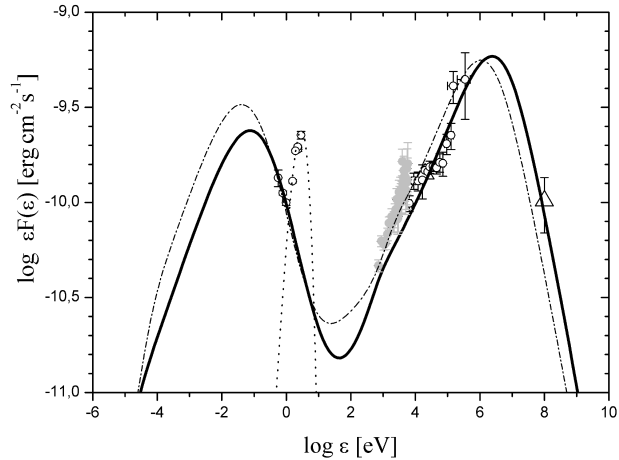}
\end{center}
  \caption{Model of the SED of 3C 273 for the first (dot-dashed line) and second (solid
  line) week of the multifrequency campaign. The data of the second week are shown. Triangle represents the AGILE data. The light grey data refer to
  XRT observations, performed during the third week [Adapted from Pacciani et
  al.~2009].} 
\label{figure9}
 \end{figure}

We organized a 3-week multifrequency campaign between mid-December 2007 and January 2008 on 3C 273
involving REM, RXTE, INTEGRAL, \textit{Swift}, and AGILE, with the aim of
studying the correlated variability in the different energy ranges and of building a time-resolved SED for each of the 3 weeks from near-IR
to $\gamma$ rays. During this campaign, whose results are reported in
Pacciani et al.~(2009), the source was detected in a high state in X-rays, with a 5--100
keV flux a factor of $\sim$3 higher than the typical value in historical
observations (Courvoisier et al. 2003), whereas it was detected in $\gamma$ rays
only in the second week, with an average flux of F$_{\rm E>100 \, MeV}$ = (33
$\pm$ 11)$\times$10$^{-8}$ ph cm$^{-2}$ s$^{-1}$. The simultaneous light
curves from near-IR to $\gamma$ rays do not show any strong correlation,
except for an indication of anti-correlated variability between X-rays and
$\gamma$ rays: all the soft and hard X-ray measurements show a decreasing
trend at the time of the AGILE detection in $\gamma$ rays in the second week. The SED is well represented by a leptonic model where the soft
X-ray emission is produced by the combination of SSC and EC models, while the
hard X-ray and $\gamma$-ray emission is due to ECD (Fig.~\ref{figure9}). The spectral
variability between the first and the second week is consistent with the
acceleration episode of the electron population responsible for the
synchrotron emission, leading to a shift of the IC peak towards higher
energies. A possible shift of the IC peak was proposed when comparing
the June 1991 campaign with the OSSE observation in September 1994
(McNaron-Brown et al.~1997). Our multifrequency observation and modeling (see
Pacciani et al.~2009) suggests that this behaviour could be a more general feature of this source, happening on
shorter timescales. Finally, considering the weak X-ray flux in the second
week of the campaign we investigated the presence of a Seyfert-like disk
reflection hump at $\sim$ 20--60 keV. The wide band spectral data from all the
instruments onboard INTEGRAL shows that the jet emission alone does not
describe perfectly the energy spectrum and a reflection hump improves slightly
the X-ray spectral modeling.    

\subsection{TeV blazars: Mrk 421 and W Comae}

With the advent of the latest generation of Imaging Atmospheric Cherenkov
Telescopes (IACTs) the number of sources detected in the TeV energy regime has
significantly increased. While the majority of TeV sources are Galactic, so
far 31 AGNs have been detected; only 8 of them were detected by
EGRET. Most of these sources were discovered at TeV
energies only by the new generation of IACTs, therefore the number of TeV blazars
detected contemporaneously at MeV--GeV and TeV energy bands is still very
low. Thus, until now multiwavelength campaign have been
largely unable to probe information on this part of the electromagnetic
spectrum. 

With the launch of two new $\gamma$-ray satellites, AGILE and \textit{Fermi}, the
gap in the MeV--GeV domain has been closed giving the possibility to remove the degeneracies in
the modeling of the SEDs of these objects. In fact, simultaneous observations
from MeV to TeV, where most of the energy of blazars is emitted, could provide
important information on the physics underlying the emission from these
objects. As first examples of the synergy between $\gamma$-ray satellites and
IACTs, multiwavelength campaigns including simultaneous AGILE, MAGIC, and VERITAS observations of Mrk 421 and W Comae were performed in June 2008.

\begin{figure*}[!hhh]
\begin{center}
\includegraphics[width=7.0cm]{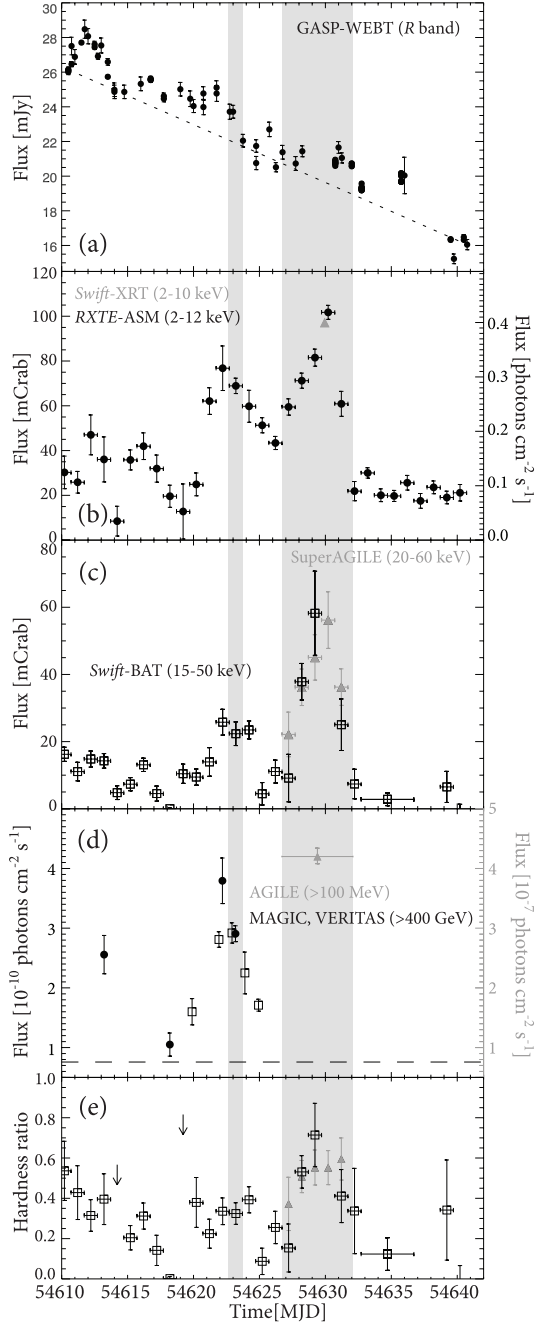}
\end{center}
 \caption{\footnotesize{\textbf{a)} $R$-band optical light curve from
GASP-WEBT (24 May--23 June 2008); \textbf{b)} ASM (2--12 keV) light
   curve and XRT (2--10 keV) flux (grey triangle);
   \textbf{c)} SuperAGILE (20--60 keV, grey triangles; 1 Crab $= 0.20 $ ph cm$^{-2}$ s$^{-1}$)
   and BAT (15--50 keV, empty black squares; 1 Crab $= 0.29$ ph cm$^{-2}$
   s$^{-1}$); \textbf{d)} MAGIC and VERITAS (E $>$ 400 GeV, empty black squares and black circles, respectively), the Crab
   flux at E $> 400$ GeV (horizontal dashed line), AGILE (E $>$ 100 MeV, grey triangle); \textbf{e)} the hardness ratio computed by using the
  SuperAGILE and ASM data for each day. The dashed vertical lines mark the
  period of the TeV flare and the GeV flare, respectively [Adapted from Donnarumma et al.~2009a]}} 
\label{figure10} 
\end{figure*}

W Comae was the first IBL object to be detected at very high energy (VHE; Acciari et al.~2008). It was discovered at TeV energies by VERITAS during observations
carried out over January--April 2008, with an integrated flux of 9$\%$ of the
Crab flux during a 4 days flare in mid-March. On 8 June 2008, VERITAS
announced the detection of a second TeV flare from W Comae (Swordy 2008), with
a three times higher flux with respect to the previous one. About 24 hours later, AGILE re-pointed towards the source and
detected it with a flux of (90 $\pm$ 32)$\times$10$^{-8}$ ph
cm$^{-2}$ s$^{-1}$ for E $>$ 100 MeV, roughly a factor of 1.5 larger than the highest flux
observed by EGRET (see Verrecchia et al.~2008). A multiwavelength campaign that included \textit{Swift}, XMM-\textit{Newton}, and GASP
observations was triggered, covering the entire electromagnetic spectrum
from radio to TeV. Acciari et al.~(2009) modeled the resulting SED during the
VHE $\gamma$-ray flare by means of a simple leptonic SSC model, but the wide separation of the two peaks in the SED
requires a low ratio of the magnetic field to the electron energy density
($\epsilon_B = 2.3\times 10^{-3}$), far from the equipartition. On the contrary, the SSC+EC model returns magnetic field parameters closer
to equipartition, providing a satisfactory description of
the broadband SED. The external radiation field could be
produced by a torus, whose emission peaks at 1.5 $\times$ 10$^{14}$ Hz, that
also explains the slight near-IR bump observed in the SED of W Comae. This bump
could also be due to the host galaxy and future observations of variability of
the IR component or very high-resolution imaging are required to break this degeneracy.

During the ToO towards W Comae, AGILE also detected the HBL object Mrk 421, in
both hard X-rays and $\gamma$ rays. SuperAGILE detected a fast increase of flux
from Mrk 421 up to 40 mCrab in the 15--50 energy band, about a factor of 10 higher
than its typical flux in quiescence (Costa et al.~2008), reaching about 55
mCrab on 13 June 2008. This observation was followed by the detection in
$\gamma$ rays by GRID with a flux, F$_{ \rm E>100\, MeV}$ =
(42 $\pm$ 13)$\times$10$^{-8}$ ph cm$^{-2}$ s$^{-1}$, about a factor
of 3 higher than the average EGRET value, and $\sim$1.5 higher, but still
consistent, with its maximum. Two strong flares at TeV energies were detected also by the
ARGO-YBJ detector between June 3 and 15 (Aielli et al.~2010). 
An extensive multiwavelength campaign from optical to TeV energy bands was
organized with the participation of WEBT, \textit{Swift}, RXTE, AGILE, MAGIC,
and VERITAS, as reported in detail in Donnarumma et al.~(2009a). 
SuperAGILE, RXTE/ASM and \textit{Swift}/BAT show a clear
correlated flaring structure between soft and hard X-rays with a high
flux/amplitude variability in hard X-rays (Fig.~\ref{figure10}). Hints of the same flaring activity is also detected in optical band
by GASP-WEBT, with an overall decreasing trend with superimposed spikes of
emission that show variations of the order of 10$\%$--20$\%$ on timescales of few
days. Moreover, \textit{Swift}/XRT observed the source at the highest 2--10 keV flux
  ever observed ($\sim$2.6 $\times$ 10$^{-9}$ erg cm$^{-2}$ s$^{-1}$), with
  the synchrotron peak at $\sim$3 keV, showing a
  shift with respect to the typical values of 0.5--1.0 keV. VERITAS and MAGIC
  observed the source on 6--8 June 2008 in a bright
  state at TeV energies, well correlated with the simultaneous peak in X-rays.
The SED of Mrk 421 can be interpreted
within the framework of the SSC model in terms of a rapid acceleration of
leptons in the jet, in accordance also with the X-ray and VHE correlation. 
However, the different behaviour at optical and X-rays could suggest an
alternative more complex scenario, in which the optical and X-ray radiation could be produced in different regions
of a helical jet, with the inner jet region that produces the X-rays and is
partially transparent to the optical radiation, whereas the outer region
produces only the lower-frequency emission. This behaviour could be explained
by a geometrical model, in which the emitting plasma flows along a rotating
helical path (see e.g.~Villata $\&$ Raiteri 1999).

\subsection{PKS 0537--441}

PKS 0537--441 is a bright and variable emitter at all frequencies from radio to
$\gamma$ rays, showing a SED typical of the LBLs. While BL Lacs usually
exhibit optical featureless spectra, in the optical/UV spectrum of this source
 strong broad emission lines of Mg II,
Ly$\alpha$ and C IV were observed (Peterson et al.~1976, Pian et al.~2002). Its relatively high
redshift (z = 0.896) and the presence of broad emission lines in its spectrum, similarly to FSRQs, make this source peculiar, and
possibly bridging the gap between BL Lacs and FSRQs.

Between 15 September and 3 October 2008 \textit{Fermi}-LAT
detected an increase of the $\gamma$-ray activity from the source (Tosti 2008), and this alert triggered multiwavelength observations by
REM, \textit{Swift}, and AGILE.
During the period 10--17 October 2008, PKS 0537--441 was detected by AGILE
with an average $\gamma$-ray flux of F$_{\rm E>100\,MeV}$ = (42 $\pm$ 11) $\times$
10$^{-8}$ ph cm$^{-2}$ s$^{-1}$ and a significant rise of the activity of
the source in the second half of the observing period, as reported in Pucella et al.~(2010). REM and \textit{Swift}/XRT detected the source in
near-infrared/optical and X-rays during a relatively low and intermediate
activity states, respectively, with no signs of evident variability in the
different observations. Instead, \textit{Swift}/UVOT detected an increase between
the first and the second part of the observing period, smaller than in
$\gamma$ rays. The SED of PKS 0537--441 in mid-October 2008 seems to require two
SSC components to be modeled, to account for the near-infrared/optical
bump, the X-ray data, and the $\gamma$-ray flux level observed by
AGILE. An alternative model where the optical bump is produced mainly by the
accretion disk emission with a relatively high luminosity is also proposed; in
this case a subsequent significant contribution of the EC scattering of both
direct disk radiation and photons reprocessed by the BLR dominates the
$\gamma$-ray band. In fact,
even if BL Lacs are usually associated with low photon density ambient and therefore the
EC contribution is negligible for this objects, past observations of
broad emission lines in the optical/UV spectrum of PKS 0537$-$441 (Pian et
al.~2002) suggests that the contribution of the EC could be important for
the high-energy emission of this object, at least during high activity
states. 

\section{Concluding remarks}

We presented a review of the results on multiwavelength studies of the
brightest blazars detected in $\gamma$ rays by AGILE in the first three years
of operation, together with the simultaneous observations collected from radio
to TeV energy bands of these objects. Based on the modeling of the SED of BL Lacs and FSRQs and the
  study of correlated variability at different frequencies, we found that the SSC and EC models are good approximation for describing the high activity states
of the two populations of blazars, respectively. By contrast, when detailed observations
are available, the fit of SEDs and the correlation studies require more complex scenarios beyond the standard emission
models. Some examples are: i) the presence of two SSC components or the
additional contribution of EC emission in BL Lacs; ii) some evidence in favour
of an helical structure of the jet in blazars; iii) in a few FSRQs a possible
contribution in the $\gamma$-ray band from the inverse Compton scattering of
seed photons from the accretion disk, the dusty torus or the hot corona. Moreover, in some cases, Seyfert-like features such as the little and big blue bumps,
the soft X-ray excess, and the Compton reflection component are detected in the broad band spectrum of FSRQs, not only
during low activity states. Finally, the long-term monitoring of these blazars
allowed us to observe complex correlations between the emission at different
frequencies and flaring episodes with extreme energetics in both BL Lacs and FSRQs.

To conclude, blazars constitute a very intriguing class of objects that show
a variety of peculiar behaviours that for over thirty years draw the attention of
astrophysicists around the world. However, early on the investigation of
blazars were made difficult by the impossibility to obtain detailed
observation of these objects over the entire electromagnetic spectrum. In
particular, the $\gamma$-ray domain remained inaccessible for over 10 years
after the end of the EGRET experiment, depriving us of important information
essential to understand the emission mechanisms at work. At last, with the
two $\gamma$-ray satellites AGILE and \textit{Fermi} in orbit, a new
window is now opened, not only for the observations in
$\gamma$ rays but also for further coordinated investigations over the whole
electromagnetic spectrum. 
It will allow us to expand the number of sources studied and the amount of
information on them, shedding light on most of the mysteries of this exciting class of objects.

\section*{acknowledgements}

F. D'Ammando would like to thank the organizers of the E-11 Event,
L. Foschini and G. Tosti, for having organized such an excellent and fruitful meeting.
The AGILE Mission is funded by the ASI with scientific and programmatic
participation by the Italian Institute of Astrophysics (INAF) and the Italian Institute of Nuclear
Physics (INFN). We thank the GASP-WEBT Collaboration for providing the data
presented here. This investigation was carried out with partial support under
ASI Contract No.~I/089/06/1.

\end{document}